\documentclass[reprint,prl,superscriptaddress,longbibliography]{revtex4-2}
\usepackage{graphicx}
\usepackage{dcolumn}
\usepackage{bm}
\usepackage[mathlines]{lineno}
\usepackage{hyperref}
\hypersetup{colorlinks=true,linkcolor=red,citecolor=blue,urlcolor=magenta}
\usepackage{multibib}
\usepackage{braket}
\usepackage{color}
\usepackage{times}
\usepackage{amssymb}
\usepackage{amsmath}
\usepackage{epsfig}
\usepackage{epstopdf}
\usepackage{dsfont}
\usepackage{subfigure}
\usepackage{verbatim}
\usepackage{float}
\usepackage{xcolor}

\begin{document}
	
\title{Non-Hermitian Sensing via a Divergent Quantum Metric}
\author{Teng Liu}
\affiliation{School of Physics and Astronomy, Sun Yat-sen University, Zhuhai 519082, China}
\author{Xiaohang Zhang}
\affiliation{School of Physics and Astronomy, Sun Yat-sen University, Zhuhai 519082, China}
\author{Jiawei Zhang}
\email{zhangjw286@mail.sysu.edu.cn}
\affiliation{School of Physics and Astronomy, Sun Yat-sen University, Zhuhai 519082, China}
\affiliation{Guangdong Provincial Key Laboratory of Quantum Metrology and Sensing, Sun Yat-sen University, Zhuhai 519082, China}
\author{Le Luo}
\email{luole5@mail.sysu.edu.cn}
\affiliation{School of Physics and Astronomy, Sun Yat-sen University, Zhuhai 519082, China}
\affiliation{Guangdong Provincial Key Laboratory of Quantum Metrology and Sensing, Sun Yat-sen University, Zhuhai 519082, China}
\affiliation{Quantum Science Center of Guangdong-Hong Kong-Macao Greater Bay Area, Shenzhen 518048, China}
\affiliation{Shenzhen Research Institute of Sun Yat-Sen University, Shenzhen 518057, China}
\affiliation{State Key Laboratory of Optoelectronic Materials and Technologies, Sun Yat-sen University, Guangzhou 510275, China}

\begin{abstract}
The quantum metric, a geometric measure of state-space distance, has recently attracted growing attention for capturing anomalous state responses to parameter variations. Especially in non-Hermitian systems, the quantum metric has been observed to diverge when the eigenstates coalesce, a phenomenon identified as a remarkable resource for sensing. Here, by exploiting this divergence, we establish a non-Hermitian sensing scheme that leverages enhanced transient dynamics to provide a geometric gain for amplifying external field signals. We confirm the critical enhancement in the Fisher information using a trapped-ion \textsuperscript{171}Yb\textsuperscript{+} platform and demonstrate superior noise robustness over conventional eigenvalue-splitting–based non-Hermitian schemes by evaluating the minimum detectable signal. Moreover, this scheme can be naturally combined with non-Hermitian topological dynamics, revealing a unique unidirectional sensing response, which indicates its potential for directional signal discrimination. Our work establishes a new paradigm for sensing in open quantum systems through critical quantum geometry and opens a route toward robust topological quantum sensing.
\end{abstract}

\maketitle
Quantum geometry, encompassing both the Berry curvature and the quantum metric, underpins a wide range of quantum phenomena, from topological states to nonlinear responses~\cite{berry1984quantal, thouless1982quantized, xiao2010berry, nagaosa2010anomalous, torma2023essay,chang2013experimental}. While Berry curvature has long been central to understanding topology and transport, the quantum metric, which quantifies the intrinsic distance between quantum states~\cite{yu2022quantum, sala2025quantum}, is increasingly recognized as a key player in phenomena beyond the adiabatic regime, such as superfluid weight~\cite{peotta2015superfluidity, julku2016geometric} and nonlinear transport in the anomalous Hall effect ~\cite{kolodrubetz2017geometry, kaplan2024unification, wang2023quantum, gao2023quantum}. Formally, these two quantities are unified in the quantum geometric tensor (QGT) \cite{provost1980riemannian}. Nevertheless, in conventional Hermitian settings, the quantum metric often appears as a subleading correction rather than a dominant physical resource. In contrast, in non-Hermitian (NH) systems, especially near exceptional points (EPs) where eigenvalues and eigenstates coalesce, the quantum metric takes on a far more decisive role~\cite{ibanez2014adiabaticity,wang2018non,solnyshkov2021quantum, tzeng2021hunting}. It diverges hypersingularly, strongly influencing non-adiabatic dynamics and enabling phenomena such as chiral state transfer and topological tunneling~\cite{lu2025dynamical, milburn2015general, doppler2016dynamically, liu2021dynamically, zhang2019dynamically, nasari2022observation}.

Recent studies show that the divergence of the quantum metric near EPs can couple directly to parametric modulations~\cite{liao2021experimental}, suggesting it may serve as a powerful and intrinsic amplifier for sensing. While existing EP-based NH sensing schemes are predominantly characterized by high eigenvalue-splitting sensitivity to parameter variations~\cite{heiss2004exceptional, guo2009observation, kato2013perturbation, fleury2015invisible, chen2017exceptional, hodaei2017enhanced, zhang2018phonon, xiao2019enhanced, wu2019observation, liu2020gain, yu2020experimental, ding2021experimental, wu2021high, ding2022non, shen2023gain,lau2018fundamental, wang2020petermann, liu2023petermann, ding2023fundamental}, they typically rely on fine-tuned signal design and come at the cost of elevated noise and dynamical instability. By contrast, harnessing the quantum metric opens a distinct and more fundamental path: robust, eigenstate-related geometric responses replace spectral singularities. The enhancement is rooted in the geometric background near EPs and, importantly, naturally integrates NH topology into sensing, offering critical sensitivity enhancement without the trade-offs of eigenvalue-based methods. 
However, a sensing framework with explicit critical effects based on the quantum metric is still lacking.

In this work, we develop and demonstrate an NH sensing framework based on the divergent quantum metric. We derive an explicit relation linking this divergence to the Hamiltonian's phase change rate, enabling dynamically amplified sensing. Experimentally, we use a trapped‑ion ($^{171}\text{Yb}^{+}$) platform that realizes a parity–time ($\mathcal{PT}$)‑symmetric Hamiltonian with established NH dynamics~\cite{lu2025dynamical}. By observing the transient tunneling dynamics near an EP, we quantify the sensitivity enhancement via Fisher information, confirming its sharp rise near the EP. Furthermore, by evaluating the minimum detectable signal, we demonstrate significantly superior noise robustness over conventional eigenvalue‑based schemes. We also reveal a unique unidirectional sensing response from chiral tunneling, effectively embedding NH topological properties into the measurement process. This joint mechanism establishes a practical pathway toward NH vector sensing.

\textit{Non-Hermitian quantum geometric tensor (NH-QGT).}---The sensing scheme is implemented in an NH $\mathcal{PT}$-symmetric atomic system with $H=i\gamma\sigma_z+J\sigma_x$~\cite{bian2023quantum,lu2024experimental,lu2024realizing,lu2025dynamical,peng2025observation,li2019observation}, where $\sigma_z $ and $\sigma_x$ are Pauli matrices, and $J$ is the Rabi strength. The sensing process is constructed in a dynamic way by incorporating a modulated Rabi strength $J(t)$ and a time-varying coupling phase $\Phi(t)$ into the Hamiltonian, giving the total parameter-dependent Hamiltonian
\begin{equation}
	H_{\mathcal{PT}}[J(t), \Phi(t)] =
	\begin{pmatrix}
		i\gamma & J(t) e^{i \Phi(t)} \\
		J(t) e^{-i \Phi(t)} & -i\gamma
	\end{pmatrix}.
	\label{HPT}
\end{equation}

\begin{figure}
	\centering
	\includegraphics[width=0.47\textwidth]{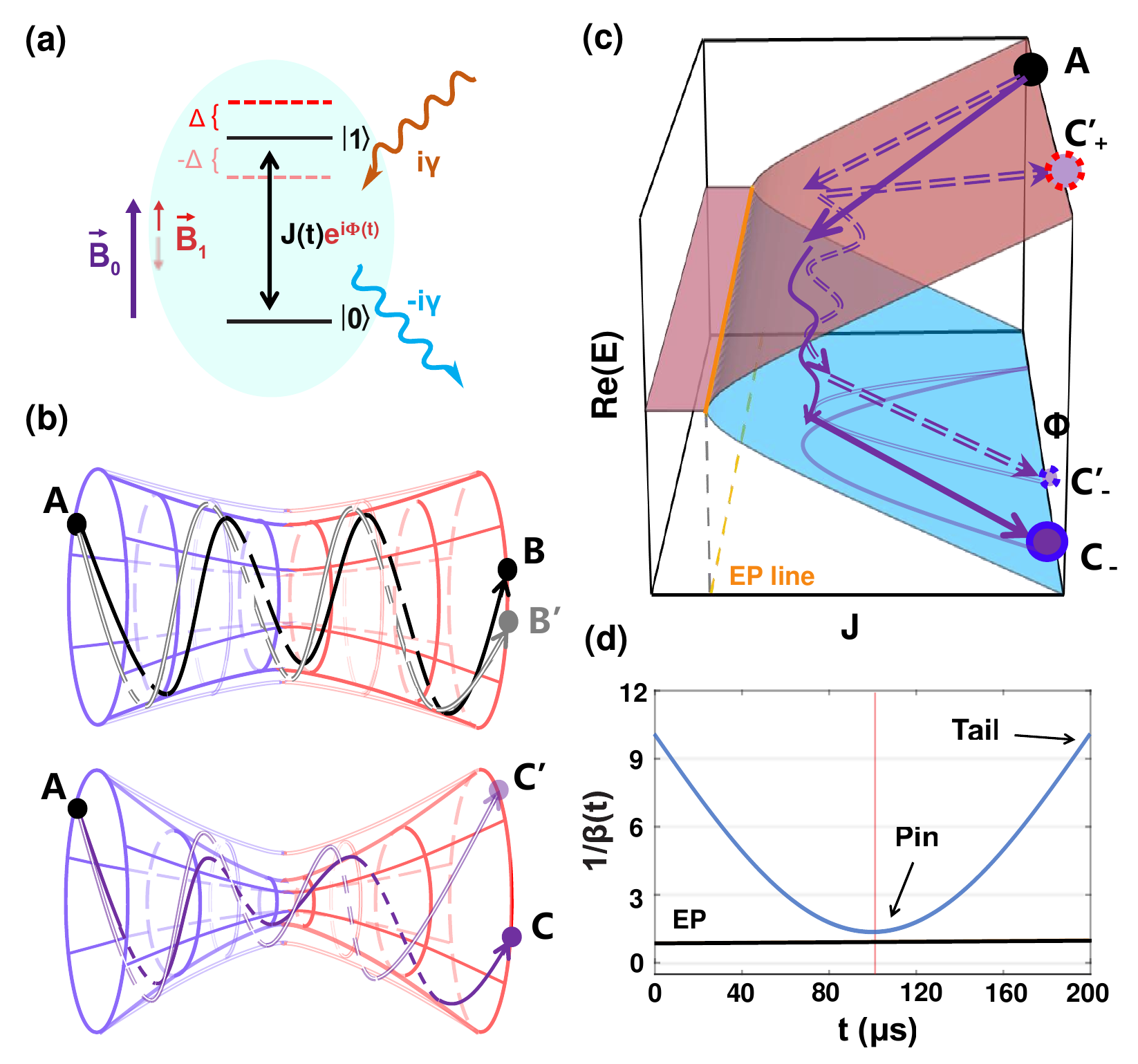}
	\caption{Schematics of non-Hermitian sensing via a divergent quantum metric. 
		(a) The energy levels of the non-Hermitian Hamiltonian $H_{\mathcal{PT}}(t)$. The shift of energy levels $\Delta$ induced by the external field being measured determines the variation of the phase $\dot{\Phi}$. (b) Conceptual figure illustrating quantum geodesic deviation driven by a divergent quantum metric. The state space is shown as a manifold, with blue and red surfaces denoting gain and loss regions governed by the imaginary part of the intraband Berry connection. The circle \(A\) marks the initial state. Upper (lower) panel: far from (near) EPs, the quantum-metric variation is suppressed (enhanced). Solid and hollow black (purple) lines indicate two state trajectories evolved under slightly different \(\dot{\Phi}\), terminating at \(B\) and \(B'\) (\(C\) and \(C'\)) after dynamic sensing. The separation between the final states visualizes quantum geodesic deviation, where the divergent quantum metric amplifies small differences in \(\dot{\Phi}\).
		(c) Schematics of non-Hermitian topological tunneling. Red (blue) branches show the real part of eigenenergies \(E\) with positive (negative) values. \(J\) is time-modulated to scan toward and then away from the EP line, as traced by the paths in the lower \(J\)–\(\Phi\) plane. From the initial state \(A\), dashed (solid) arrows denote process under small (large) \(\dot{\Phi}\). The final states \(C\) and \(C'\) exhibit different populations in both branches (\(C_{-}\), \(C'_{+}\), \(C'_{-}\)), indicated by circle sizes. Near EPs, metric enhancement amplifies a minute signal difference into a pronounced separation of the final states after tunneling.
		(d) Modulation function \(\beta(t)\) during dynamic sensing. The system starts at the ``tail'' far from the EP line (orange dashed line, \(\beta=1\)), scans to the ``pin'' near the EP line (red line, \(\beta_p=0.75\)), and then returns to the ``tail''.
	}
	\label{one}
\end{figure}

For a constant target signal to be sensed (e.g., an electromagnetic field), the induced level shift can be quantified as a detuning $\Delta$, which is incorporated into the Hamiltonian, yielding the phase $\Phi(t) = \Delta t$, as shown in Fig.~\ref{one}(a). Consequently, the susceptibility of the system to the target signal is converted into susceptibility to variations of a parameter $\Phi(t)$ in the Hamiltonian, as governed by $\Delta = \dot{\Phi}(t)$. More details are shown in Supplementary Materials~(SM)~\cite{SupplementalMaterial}~Sec.~I. 

Within the $J$--$\Phi$ parameter manifold, the system's susceptibility to infinitesimal parameter variations can be quantified geometrically when approaching EPs, where the quantum
metric is expected to diverge due to eigenstate coalescence. To characterize this effect, we introduce the NH-QGT. Following our previous work on NH trapped-ion qubits~\cite{bian2023quantum,lu2024experimental,lu2025dynamical,lu2024realizing}, we adopt the $\mathcal{CPT}$ inner product $\langle a \mid b \rangle^{\mathcal{CPT}}$~\cite{bender2007faster} (the subscript $\mathcal{CPT}$ is omitted hereafter unless stated otherwise). We then obtain the intraband and interband Berry connections, $\mathcal{A}_{\nu}^{nn}=i\langle \phi_n| \partial_\nu|\phi_n\rangle$ and $\mathcal{A}_{\nu}^{mn}=i\langle \phi_m| \partial_\nu|\phi_n\rangle$, respectively (see SM Sec.~II for details). The NH-QGT is defined as $\mathcal{T}_{\mu \nu}=\sum_{m \neq n} \mathcal{A}_{\mu}^{nm} \mathcal{A}_{\nu}^{mn}$, with the corresponding quantum metric tensor given by $g_{\mu \nu}\equiv\operatorname{Re} \mathcal{T}_{\mu \nu}$. The quantum metric quantifies the infinitesimal distance $ds^2 = g_{\mu\nu} d\lambda^\mu d\lambda^\nu$ between neighboring quantum states under variations of parameters $d\lambda^\mu$ and $d\lambda^\nu$. Its elements set how strongly the state changes under parameter variations, providing a geometric measure of the sensitivity in sensing~\cite{bleu2018effective,song2024fast,kang2025measurements,braunstein1994statistical}.

\textit{NH-QGT enabled dynamical EP sensing.}---For the NH Hamiltonian $H_{\mathcal{PT}}[J(t), \Phi(t)]$ and the corresponding $J$--$\Phi$ parameter manifold, the distance between neighboring quantum states reduces to 
\begin{equation}
d s^2 = g_{\Phi \Phi}(t) (d \Phi)^2,
\end{equation} 
where $g_{\Phi\Phi}(t) = 1/[4(1-\beta(t)^2)]$ (see SM~Sec.~II). Here, $\beta(t) = \gamma/J(t)$ is the effective dissipation. When $\beta(t) \rightarrow 1$ (close to the EPs), $g_{\Phi \Phi}$ diverges. Consequently, even a small parameter change $d\Phi$ can produce a significant distance between the adjacent states, meaning a strongly enhanced response to perturbations in $\Phi$.  As shown in Fig.~\ref{one}(b), the quantum metric remains small away from the EP (upper panel), resulting in near-parallel evolution trajectories. Near the EP (lower panel), its divergence sharply amplifies the geometric distance between trajectories under parameter variation—an effect we can term “quantum geodesic deviation” in analogy to its classical counterpart. This geometric amplification enhances the readout and increases the Fisher information.

Central to the above sensing resource is to measure the transient-state tunneling near the EP as the system evolves in parameter space. This converts the geometrically enhanced susceptibility into a measurable susceptibility to the dynamical evolution of the system. In contrast to the NH tunneling in the adiabatic regime studied in Ref.~\cite{lu2025dynamical},
here we utilize the tunneling in the non-adiabatic regime to amplify the signal, where the divergent quantum metric plays a role. According to the method in Ref.~\cite{ibanez2014adiabaticity}, the adiabatic condition is modified by the quantum metric as (see SM~Sec.~II for details)
$
\frac{\xi(t)|\dot{\Phi}(t)|}{\left|E_n(t)-E_m(t)\right|} \ll 1,
\label{nc}
$
where $E_{n,m}$ are eigenvalues. There is a geometric amplification factor 
\begin{equation}
	\xi(t)=\sqrt{g_{\Phi\Phi}^{\mathcal{CPT}}(t)}=\frac{1}{2\sqrt{1-\beta(t)^2}}
\end{equation}
linearly coupling to $\dot{\Phi}(t)$. Thus, with a divergent quantum metric, the adiabatic condition is fragile; a tiny $\dot{\Phi}(t)$ (related to the sensing signal) will result in significant tunneling as shown in Fig.~\ref{one}(c).

\begin{figure}
	\centering
	\includegraphics[width=1\columnwidth]{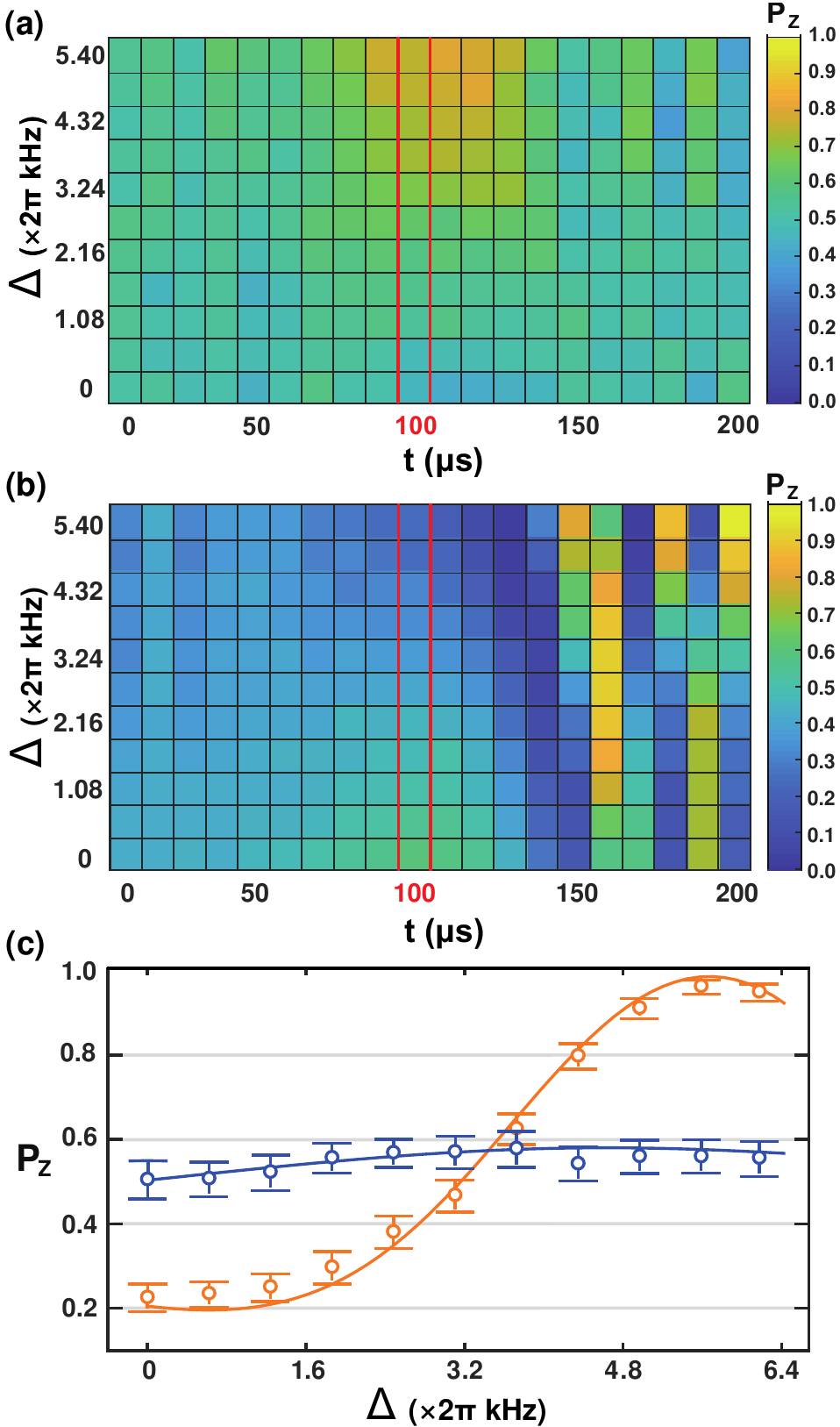}
	\caption{Experimental results of the time-dependent response of dynamic non-Hermitian sensing. 
		(a) Non-Hermitian: using the modulation parameters in Fig.~\ref{one}(a), for various detuning $\Delta$, the time evolution of $P_z$ represented by the color map. 
		(b) Hermitian: using the same modulation, while setting the dissipation to zero, the system becomes Hermitian, $P_z$ is also measured.
		(c) Comparison of the final-state $P_z(T=200\mu s)$ population with the different signal $\Delta$ being measured. The orange points represent the non-Hermitian sensing in (a), while the blue points correspond to the Hermitian case in (b). The non-Hermitian regime shows higher signal sensitivity.}
	\label{two}
\end{figure}

In practice,  we design a modulation protocol $\beta(t)$ that first approaches and then departs from the EPs along a smooth trajectory, as shown in Fig.~\ref{one}(d). During this process, the maximum value $\beta^{\rm max}(t=\tau_p)$ (where $\dot\beta(\tau_p)=0$) occurs at the position closest to the EPs, at which tunneling is most likely to occur. For instance, $\beta(t)$ can be defined as a sinusoidal function
\begin{equation}
	\beta(t) = \frac{1}{[(\beta_p^{-1} - \beta_l^{-1}) \sin\left(\pi t/T\right) + \beta_l^{-1}]}, 
	\label{modu}
\end{equation}
where $T$ is the total time, and $\beta_p=\gamma/J_p$ and $\beta_l=\gamma/J_l$ represent the ``pin” close to the EP line ($\beta=1$) and the ``tail” away from the EP line, respectively. We control the whole range of $\beta(t)$ within the $\mathcal{PT}$-symmetric regime and enhance $\xi(t)$ by tuning $\gamma$ to bring the ``pin" closer to the EP line. 
The parameter in the Hamiltonian is then determined by $J(t)=\gamma/\beta(t)$, while $\Phi(t)$ is encoded by the target signal.

The system's response is detected by measuring the final-state population after evolving the initial state under the Hamiltonian $H_{\mathcal{PT}}[J(t), \Phi(t)]$. During the evolution (a tunneling process), the wave function is expressed in the instantaneous eigenbasis $\{\ket{\phi_1(t)}, \ket{\phi_2(t)}\}$ as $\ket{\psi(t)}=\alpha_1(t)\ket{\phi_1(t)}+\alpha_2(t)\ket{\phi_2(t)}$. The dynamics of the expansion coefficients $\alpha_{1,2}(t)$ is governed by the equation of motion (EOM) $\dot{\alpha}_m(t) = i \sum_{n=1}^{2} \left[\mathcal{A}_{mn}(t) - \mathcal{H}_{mn}(t)\right] \alpha_n(t),
	\label{alpha}$ 
where $\mathcal{H}_{mn}(t)=\bra{\phi_m(t)}H_{\mathcal{PT}}[J(t), \Phi(t)]\ket{\phi_n(t)}$ and $\mathcal{A}_{mn}(t)=i\bra{\phi_m(t)}\partial_t\ket{\phi_n(t)}$ is known as the time-domain Berry connection. We prepare the initial state in one of the eigenstates (e.g., $\ket{\phi_1(0)}$). Tunneling between eigenstates is then manifested as an exchange of population between $\alpha_1(t)$ and $\alpha_2(t)$ via this EOM (see SM Sec.~III for details).

Meanwhile, this tunneling process also inherits the chiral and non-reciprocal nature of non-Hermitian topological tunneling from earlier studies~\cite{lu2025dynamical}. This stems from how the time-domain Berry connection $\mathcal{A}_{mn}(t)$ in the EOM depends on the sign of the parameter $\Phi$, i.e., positive or negative $\dot{\Phi}(t)$. The time-domain intraband Berry connection $\mathcal{A}_{nn}$ in the EOM is given by 
\begin{equation}
	\begin{split}
		\mathcal{A}_{nn}(t)&=-\frac{\dot\Phi(t)}{2}+i\,\frac{(-1)^n \beta(t) }{2 \sqrt{1-\beta(t)^2}}\text{sgn}(\dot\Phi(t))\cdot|\dot\Phi(t)|.
		\label{Ann}
	\end{split}
\end{equation}
The sign of $\text{Im} \mathcal{A}_{nn}$ defines the geometric gain (negative) or loss (positive) of the band~\cite{liang2013topological}. This geometric property determines the direction of the non-adiabatic tunneling near the EPs, studied in detail in our previous work~\cite{lu2025dynamical}, where the chiral and nonreciprocal tunneling always occurs from the loss band to the gain one. Eq.~(\ref{Ann}) shows $\text{Im} \mathcal{A}_{11}(t) = -\text{Im} \mathcal{A}_{22}(t)$, allowing for a unidirectional sensing scheme. 

The unidirectional sensing is state-dependent: (i) For the system initialized in state $n=1$,  if $\Phi$ increases (decreases) over time, i.e., $\text{sgn}(\dot\Phi(t))<0$ ($\text{sgn}(\dot\Phi(t))>0$), the condition $\text{Im} \mathcal{A}_{11}(t)>0$ ($\text{Im} \mathcal{A}_{11}(t)<0$) is satisfied. This leads to the occurrence (suppression) of tunneling, and the sensor  responds (does not respond) to the external signal. (ii) Conversely, initializing in state $n=2$ reverses the behavior.

We note that these distinctive topological properties enable the system to respond unidirectionally to signals, thereby offering potential for constructing a vector sensor (see SM Sec.~III for details).

\begin{figure}[ht!]
	\centering
	\includegraphics[width=1\columnwidth]{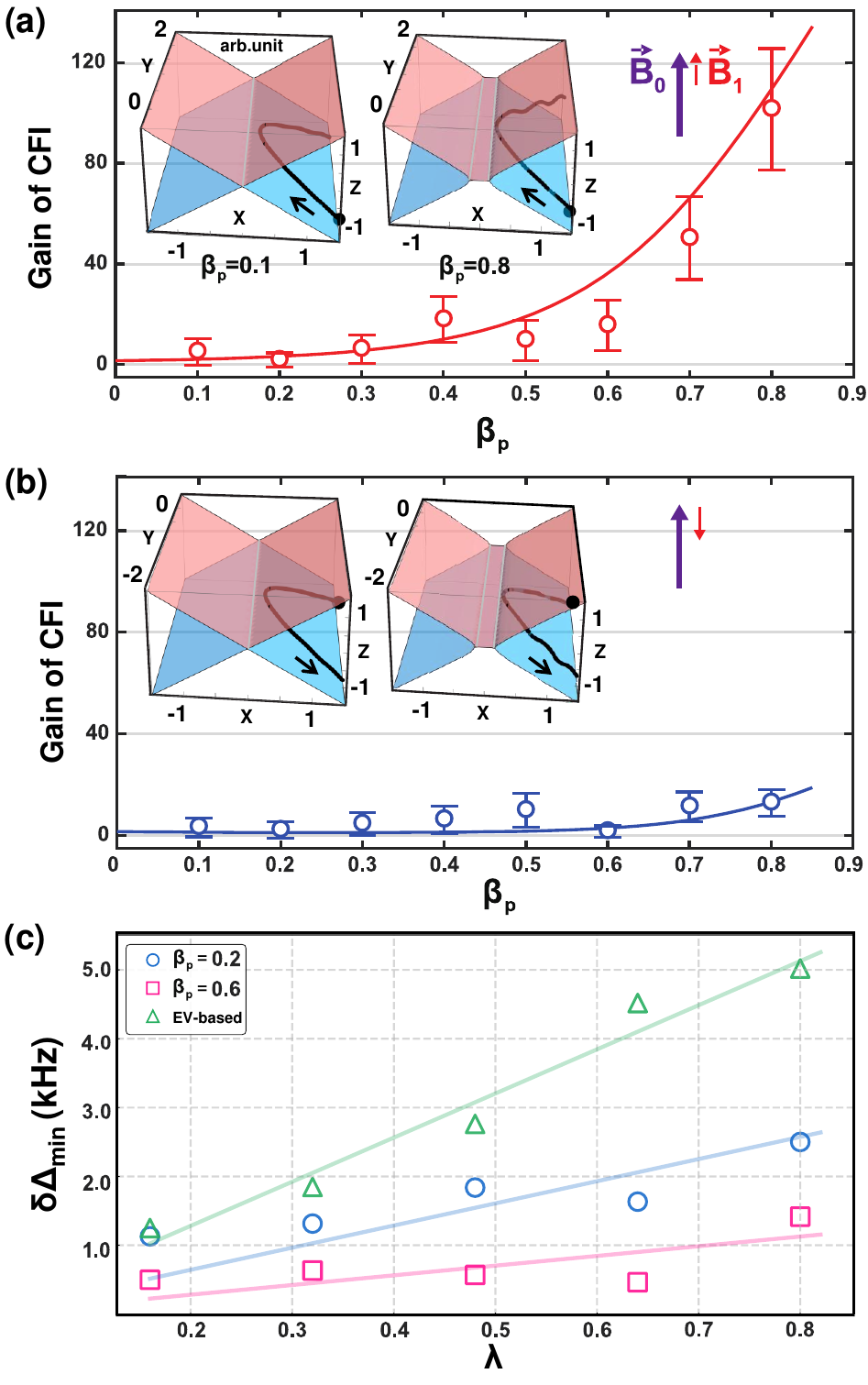}
	\caption{Experimental tests of criticality enhancement, unidirectionality, and robustness.
		(a) The state is initialized in $\ket{\phi_2}$ with $\dot{\Phi}(t) > 0$. The solid red curves represent the simulation of the gain of the CFI, and the points with error bars represent measured ones from the experimental data. For each point, the CFI obtained from the tunneling data across the detuning range $\Delta/2\pi \in (-5.9,  5.9)$ kHz. Insets: state evolution over time for $\beta_p=0.1$ and $0.8$. Axes correspond to coupling strength $J$ (X), phase $\Phi$ (Y), and energy magnitude (Z). The red (blue) surface depicts positive (negative) real eigenvalues in the $\mathcal{PT}$-symmetric regime, associated with eigenstate $\ket{\phi_1}$ ($\ket{\phi_2}$). Black circles indicate initial states, lines show evolution trajectories, and arrows mark direction. (b) The state is initialized in $\ket{\phi_2}$ with  $\dot{\Phi}(t) < 0$. 
		(c) Minimum resolvable signal $\delta\Delta_{\min}$ under different noise strengths $\lambda$. Blue circles and pink squares represent experimental results for $\beta_p = 0.2$ and $0.6$, respectively; green triangles show EV-based simulation results. Solid lines indicate corresponding linear fits. Each noise strength was tested with 15 measurements ($N_r = 15$).}
	\label{three}
\end{figure}
\textit{Experiments.}---We implement this scheme using a single trapped $^{171}$Yb$^+$ ion qubit with a $\mathcal{PT}$-symmetric Hamiltonian. Such systems have been prepared in our previous work~\cite{bian2023quantum,lu2024experimental,lu2025dynamical,lu2024realizing} and detailed in SM~Sec.~IV. Along the quantization axis of a constant magnetic field $\boldsymbol{\vec{B_0}}$, we add a magnetic field $\boldsymbol{\vec{B_1}}$ as the signal being measured. $\boldsymbol{\vec{B_1}}$ generates a detuning $|\Delta| \approx \eta |\boldsymbol{\vec{B_1}}|$, where $\eta$ is the linear Zeeman coefficient. The sign of $\Delta$ is related to the direction of vector $\boldsymbol{\vec{B_1}}$, as shown in Fig. \ref{one}(a). According to the relation $\Phi(t)=\Delta t$, the system's unidirectional response to $\text{sgn}(\dot\Phi(t))$ can thus be mapped to a unidirectional response to the direction of a vector signal. Here, we adopt the function $\beta(t)$ defined in Eq.~\ref{modu}, with the parameters set to $J_p = 4 \times 2\pi$ kHz, $J_l = 30 \times 2\pi$ kHz, and a total duration of $T = 2\tau_{p} = 200$ $\mu$s.

In the first experiment, we measure the non-adiabatic tunneling of the qubit between its NH eigenstates under various values of $\Delta$ (i.e., corresponding to different values of $\boldsymbol{\vec{B_1}}$), and obtain the dependence of the tunneling rate on the detuning signal. The initial state is prepared as $|\phi_{2}(0) \rangle$ (i.e., $|- \rangle_x$) with $\Phi(0)=0$ by applying a $\pi/2$ pulse to $\ket{0}$. Then the NH qubit evolves with the modulation function $J(t)$ in $H_{\mathcal{PT}}(t)$ with $\gamma=0.75 J_p$ (the corresponding $\beta(t)$ is presented in Fig. \ref{one}(d)), and the normalized populations of $\ket{0}$, denoted as $P_z$, are measured at various times. The experimental results are presented in Fig.~\ref{two}(b). We also apply the modulated $H_{\mathcal{PT}}(t)$ under the Hermitian conditions $\beta_p=0$ ($\gamma=0$) for comparison, shown in Fig.~\ref{two}(a).

The modulation rate of $J(t)$ in $H_{\mathcal{PT}}(t)$ satisfies the adiabatic condition $\left|\frac{\dot{J}(t)\Delta}{(\Delta^2 +{J}(t)^2)^{3/2}}\right| \ll 1$, so that, in the Hermitian case of Fig.~\ref{two}(a), the evolution remains adiabatic, and the final state preserves its initial form $|-\rangle_x$  with $P_z \approx 0.5$. It is noted that when the signal (i.e., $\Delta$) is large, e.g., around 5 kHz,  $P_z \approx 1$ near the ``pin" position, then falls back to around 0.5. This effect is due to the dynamics of the eigenstates adiabatically following the modulation of the Hamiltonian. In contrast, in the non-Hermitian case of Fig.~\ref{two}(b), non-adiabatic tunneling arises for the same modulation of $J(t)$. Meanwhile, as the signal increases, the tunneling onset position moves closer to the ``pin" position, characterized by a sharp drop in $P_z$ from around 0.5 to 0.0 and subsequent oscillations during the modulation. The final population, therefore, is highly sensitive to the value of $\Delta$, as evidenced by the oscillating colors in Fig.~\ref{two}(b) and (c). This strong dependence demonstrates the potential of utilizing this tunneling phenomenon for sensing applications.

The second experiment is to validate that the system's response to $\boldsymbol{\vec{B_1}}$ is both criticality-enhanced and unidirectional. As shown in Fig.~\ref{three}, the system initialized in $\ket{\phi_{2}(0)}$ exhibits distinctly chiral tunneling. Unidirectional tunneling occurs only when $\boldsymbol{\vec{B_1}}$ is aligned with $\boldsymbol{\vec{B_0}}$ (Fig.~\ref{three}(a)), whereas it is completely suppressed when $\boldsymbol{\vec{B_1}}$ is opposed to $\boldsymbol{\vec{B_0}}$ (Fig.~\ref{three}(b)), directly demonstrating the chiral nature of the non-Hermitian tunneling. We note that in Fig.~\ref{three}(b), the slight increase in the amplification ratio as $\beta_p$ increases originates from minor non-adiabatic oscillations caused by rapid parameter modulation, since no tunneling occurs during this process. Additionally, for the initial state $\ket{\phi_{1}(0)}$, the behavior is reversed, and the results are shown in SM~Sec.~IV. To quantify that the tunneling is criticality-enhanced near EPs, the classical Fisher information (CFI) is measured, which is defined as 
$
I(\Delta)=\left(1 / N\right) \sum_{j=1}^{N}\left[\left(d P_z / d \Delta_j\right)^2 / P_z\right],
$
where $\Delta_j$ represents the detuning in the range $(0, 5.9\times2\pi)$ kHz, and $N$ is the number of sampling points. In Fig.~\ref{three} (a) and (b), the measured gain of CFI (defined as $I_{\beta_p}/I_0$, where $I_{\beta_p}$ ($I_{0}$) indicates the CFI at $\beta_p$ ($\beta_{p}=0$)) shows a quickly divergent behavior, revealing that the system's sensitivity to probing the differential of the signal is strongly enhanced, which is controlled by the divergent quantum metric when the system approaches the EPs. We then compare the CFI obtained from experimental data and the one calculated from the simulation of the state evolution, showing nice agreement between experiments and theory based on NH-QGT.

\textit{Noise-resilient performance.}---The conventional eigenvalue (EV)-based sensing paradigm exploits the non-analytic eigenvalue, e.g., the square-root eigenfrequency splitting near a second-order EP, $\delta\omega_{\rm EP}\simeq 2\sqrt{\alpha\epsilon}$, with susceptibility $\left|\partial \delta\omega_{\rm EP}/\partial \epsilon\right|\simeq\sqrt{\alpha/\epsilon}$ , where $\epsilon$ is the signal and $\alpha$ is a system parameter \cite{hodaei2017enhanced}. 
However, the susceptibility of this scheme is highly susceptible to noise arising from control manipulation.
Our metric-enhanced protocol instead leverages the singular quantum geometry induced by eigenstates. This leads to two key features: (i) The signal is decoupled from the eigenspectrum. The susceptibility enhancement is individually governed by the local geometric background encoded in $g_{\mu\nu}$ and is tunable via the control parameter $\beta$ independently, rather than requiring the signal $\epsilon$ to be pushed to a smaller scale to enhance $\left|\partial \delta\omega_{\rm EP}/\partial \epsilon\right|$. (ii) The signal is decoupled from systemic control fluctuations. The sensing is realized through a dynamical evolution along a trajectory in parameter space. The final state signal depends on the integral effect along the path, while this process effectively averages control fluctuations such as white noise. As a result, the protocol exhibits improved robustness to control noise, a major practical limitation of EV-based EP sensors \cite{lau2018fundamental, wang2020petermann, liu2023petermann, ding2023fundamental}.

In Fig.~\ref{three}(c), we experimentally evaluate our protocol's robustness against additive white noise on the control parameter $J(t)$ and benchmark it against simulations of the EV-based EP sensor under the same noise strength. The noisy control is modeled as $J_{\mathrm{noise}}(t)=J(t)+\xi(t)$, with $\xi(t)\sim\mathcal{N}(0,\sigma^{2})$ and $\sigma=\lambda J_{p}$, where $\lambda\in(0,1)$ is the noise-strength coefficient. For the eigenfrequency-based scheme, the EP-induced splitting is $\delta\omega_{\mathrm{EP}}(\Delta)=\sqrt{2J\Delta}$; the simulation uses the same noise model, $J_{\mathrm{noise}}=J_p+\xi$. Both schemes share identical working parameters $J_{p}/(2\pi)=4~\mathrm{kHz}$ and $\Delta/(2\pi)=3.2~\mathrm{kHz}$. For each $\lambda$, we extract the minimum resolvable signal $\delta\Delta_{\min}=\sigma_{\mu}/(\sqrt{N_r}\chi)$, where $N_r$ is the number of repetitions, $\sigma_{\mu}$ is the standard deviation of the output $\mu$ over the $N$ noisy trials, and $\chi=\left|\partial\langle \mu(\Delta)\rangle/\partial\Delta\right|$ is the sensitivity at the operating point. We take $\mu(\Delta)=\delta\omega_{\mathrm{EP}}(\Delta)$ for the EV-based scheme and $\mu(\Delta)=P_z(\Delta)$ for our protocol. The results show that increasing noise leads to an approximately linear increase in the minimum resolvable signal $\delta\Delta_{\min}$ in both schemes, indicating a reduced signal-to-noise ratio (SNR). However, our protocol consistently yields a smaller $\delta\Delta_{\min}$ than the EV-based scheme, maintaining a higher SNR and demonstrating stronger robustness to noise. Meanwhile, operating closer to the EP further reduces $\delta\Delta_{\min}$. While this is mainly due to the enhanced susceptibility, it also indicates that the noise suppression remains effective even at criticality (see SM Sec.~IV). This differs significantly from previous non-Hermitian sensing schemes, where noise from the nonorthogonality of eigenmodes (Petermann-factor noise \cite{wang2020petermann,liu2023petermann}) is amplified with the signal, preventing any net SNR enhancement.

\textit{Discussion.}---We demonstrate an NH-QGT-enabled dynamical EP sensing scheme that exploits the divergence of the quantum metric near EPs. This metric divergence effectively stretches the local parameter space, converting minute parameter variations into readily detectable signals. Owing to its dynamical, trajectory-based readout, our NH-QGT approach is distinct from two existing EP-sensing strategies: first, conventional eigenvalue-splitting (spectral-singularity) sensors, where noise can undermine the practical SNR despite the enhanced spectral susceptibility~\cite{chu2020quantum, mcdonald2020exponentially}; and second, spectrum-bypassing approaches that improve robustness only by trading away the extreme susceptibility near the EP. ~\cite{xiao2024non}. 
We further show that the protocol can exhibit a unidirectional signal response, enabling one-dimensional discrimination of signal direction and suggesting a route toward a complete non-Hermitian vector-sensing framework. This work establishes a sensing paradigm rooted in eigenvector-based quantum geometry in NH systems. Looking ahead, extending this approach to many-body platforms and to higher-order exceptional points offers promising directions for future study.

The authors acknowledge Dr. Xinxin Rao and Mingshen Li for experimental support, and thank Dr. Konghao Sun and Xiaodong Tan for helpful discussions. This work is supported by the National Key Research and Development Program of China, ``Gravitational Wave Detection'' Special Project under Grant No. 2022YFC2204402, the National Natural Science Foundation of China under Grants No. 12074439 and No. 12304315, Guangdong Provincial Quantum Science Strategic Initiative under Grants No. GDZX2203001 and No. GDZX2303003, and Shenzhen Science and Technology Program under Grant No. JCYJ20220818102003006.

\end{document}